\documentclass[10pt,oneside,english,onecolumn,a4paper]{article}
\usepackage{graphicx}
\usepackage[utf8]{inputenc}
\usepackage{mathptmx}
\usepackage{tabularx}
\usepackage{ragged2e}
\usepackage[singlelinecheck=false]{caption}
\usepackage[T1]{fontenc}
\usepackage[numbers]{natbib}
\usepackage{lscape}
\usepackage{url}
\usepackage{amsmath}
\RequirePackage[blocks]{authblk}
\usepackage{babel}
\usepackage{balance}
\usepackage[nolist]{acronym}
\makeatletter
\let\ps@plain\ps@empty
\def\@xivpt{14bp}

\def\@sect#1#2#3#4#5#6[#7]#8{%
	\ifnum #2>\c@secnumdepth
	\let\@svsec\@empty
	\else
	\refstepcounter{#1}%
	\protected@edef\@svsec{%
		\ifnum #2<4
		\hb@xt@10mm{\csname the#1\endcsname}\relax
		\else
		\hb@xt@12mm{\csname the#1\endcsname}\relax
		\fi}%
	\fi
	\@tempskipa #5\relax
	\ifdim \@tempskipa>\z@
	\begingroup
	#6{%
		\@hangfrom{\hskip #3\relax\@svsec}%
		\interlinepenalty \@M #8\@@par}%
	\endgroup
	\csname #1mark\endcsname{#7}%
	\addcontentsline{toc}{#1}{%
		\ifnum #2>\c@secnumdepth \else
		\protect\numberline{\csname the#1\endcsname}%
		\fi
		#7}%
	\else
	\def\@svsechd{%
		#6{\hskip #3\relax
			\@svsec #8}%
		\csname #1mark\endcsname{#7}%
		\addcontentsline{toc}{#1}{%
			\ifnum #2>\c@secnumdepth \else
			\protect\numberline{\csname the#1\endcsname}%
			\fi
			#7}}%
	\fi
	\@xsect{#5}}
\renewcommand\LARGE{\@setfontsize\LARGE{16}{20}}
\def\abstract#1{\def\@abstract{#1}}
\def\abstractEn#1{\def\@abstractEn{#1}}
\def\titleEn#1{\def\@titleEn{#1}}
\def\@maketitle{%
	\newpage
	\null
	\let \footnote \thanks
	{\LARGE\bfseries\RaggedRight \@titleEn \par}%
	\vskip 1\baselineskip%
	{\normalsize
		\@author\par}%
	\vskip \baselineskip%
	{\section*{Abstract}
		\@abstractEn}%
	\par
	\vskip 3\baselineskip}

\renewcommand\section{\@startsection {section}{1}{\z@}%
	{-3.5ex \@plus -1ex \@minus -.2ex}%
	{\baselineskip}%
	{\normalfont\Large\bfseries\RaggedRight}}
\renewcommand\subsection{\@startsection{subsection}{2}{\z@}%
	{\baselineskip}%
	{1ex}%
	{\normalfont\large\bfseries\RaggedRight}}
\renewcommand\subsubsection{\@startsection{subsubsection}{3}{\z@}%
	{1\baselineskip}%
	{3bp}%
	{\normalfont\normalsize\bfseries\RaggedRight}}
\renewcommand\paragraph{\@startsection{paragraph}{4}{\z@}%
	{1\baselineskip\@plus1ex \@minus.2ex}%
	{3bp}%
	{\normalfont\normalsize\RaggedRight}}
\renewcommand\subparagraph{\@startsection{subparagraph}{5}{\parindent}%
	{3.25ex \@plus1ex \@minus .2ex}%
	{-1em}%
	{\normalfont\normalsize\bfseries\RaggedRight}}
\parindent\p@
\makeatother
\DeclareCaptionLabelSeparator{enskip}{\enskip}
\usepackage{blindtext}
\usepackage{placeins}
\usepackage{changes}
\definechangesauthor[name={Hans D. Schotten}, color=red]{hs}
\definechangesauthor[name={Jacob Kochems}, color=violet]{jk}
\setremarkmarkup{(#2)}
\usepackage{siunitx}
\graphicspath{ {img/} }

\begin{document}
	\titleEn{AMMCOA - Nomadic 5G Private Networks}
	
	\author{Jacob Kochems}
	\author{Hans D. Schotten}
	
	\affil{ University of Kaiserslautern \    \{kochems, schotten\}@eit.uni-kl.de}
	
	\abstractEn{
		This paper presents ideas and concepts for interconnected off-road
		vehicles, like harvesters or tandem rollers, which span a mobile
		network by themselves to be, to some extent, independent from the
		network infrastructure.  Multiple RATs are used to cover a variety of
		use cases, including mmW technology for short range high bandwidth
		communication and ranging. The AMMCOA Project aims at providing
		connectivity in areas where the infrastructure network is either at
		long range distances or out off range entirely. In this case, for
		certain network services, like authentication of UEs, to be
		available, some of the core network functionalities have to be
		instantiated locally. This effectively turns the local network into a
		light version of a core network. In this paper we give an overview of
		the project's use cases, their requirements, the architectural idea
		of a local autonomous network (Trust Zone, 5G Island, 5G Private Networks)
		and the AMMCOA network architecture envisioned so far.}
	
	\maketitle
	
	\section{Introduction}
	The operation of agricultural and construction machines increasingly resembles complex industrial manufacturing processes. Accordingly, the requirements for efficiency, precision and safety are high. Working autonomously enables the optimization of whose processes and therefore increases productivity. Only those machines and devices employing such concepts will be competitive in the future and able to contribute to agricultural production and construction.
	These use cases exhibit usually off road characteristics.
	For those areas, digital maps are rare due to economical reasons, but high accuracy pertaining relative and absolute localization is still needed, in order to avoid straying into a neighboring field or vehicle collisions.
	A local and mobile 5G infrastructure, capable of working on its own or integrated in the mobile network infrastructure, is envisioned to enable those use cases.
	
	Project AMMCOA (Autonomous Mobile Machine Communication for Off-Road Applications) researches network solutions that are nomadic and can operate autonomously, i.e., even if no connection is available to the mobile network infrastructure.
	The intended solution should provide 5G-like network performance \cite{Afif5G1}, \cite{Afif5G2} and, in particular, meet the requirements  defined for 5G Ultra-Reliable Low Latency Communications.
	It integrates automatically in existing mobile networks and is therefore a dynamical extension of the 5G infrastructure matching current demands.
	Additionally, an integrated, high accuracy localization solution, specifically tailored to the requirements of this uses cases, will be developed.
	
	One of the use case classes this project is concerned with is the cooperative operation of agricultural machines.
	This is a common scenario in modern farming.
	To prevent duplication of effort or overdosing and to coordinate crop transfer a M2M and M2I communication as well as a high accuracy positioning method is necessary.
	
	The other use case class is concerned with the road construction scenario.
	The coordinated positioning of all the involved vehicles is the major challenge here.
	Based on movements of the paver other vehicles like tandem rollers are controlled autonomously.
	The next section features a more detailed description of the use cases.
	
	The developed solution serves as a local extension of the mobile network operator's (MNO) 5G networks.
	Its usage is therefore not just limited to the use cases outlined in this paper.
	It can also be used in scenarios where an independent and potentially temporary 5G networking between vehicles and locally placed machines is needed.
	Thus, \mbox{AMMCOA} networking concept might also become a key component of so-called 5G private networks that are currently finding interest in various vertical industries as automation, logistics, construction, and agriculture.
	
	The 5G networking solution developed in AMMCOA has - depending on the context - either to function as a stand-alone network or to serve as a local extension of the infrastructure. This  requires a new class of network architecture.
	Three related concepts are explored in this paper, their similarities and differences.
	
	The AMMCOA solution will in particular be based on so-called context-enabled radio network management. \cite{ctx1}, \cite{ctx2}, and \cite{ctx3} introduce and describe context management solutions that collect information on the networking status but on the physical and operational level as well, fusion these data sources and provide the resulting information as context information, on the one hand, to the network manegement and control and, on the other hand, to the application level.
	{\let\thefootnote\relax\footnote{This is a preprint, the full paper has been submitted to 23th VDE/ITG Conference on Mobile Communication (23.  VDE/ITG Fachtagung Mobilkommunikation), Osnabrück, May 2018}}\textbf{}
	
	\section{Use Cases and Requirements}
	\subsection{Agricultural Use Case Class}
	\begin{table*}[t]
		\centering
		\caption{Agricultural Use Cases and Requirements}\label{tab:agr-use-cases}
		\resizebox{\textwidth}{!}{%
			\begin{tabular}{|l|p{3cm}|p{5.3cm}|l|l|l|l|p{1.5cm}|}
				\hline  & & & & & & &\\
				ID      & Use Case                                & Description
				& Scope & Range & Throughput & latency & Application\\
				\hline  & & & & & & &\\
				
				1       & Transport on Roads (with reception)     & in the presence of infrastructure the vehicle can be remote controlled
				& global & N/A & \si{10}{Mbps} & \si{100}{ms} & navigation, status info\\
				
				1b      & Transport off Roads (without reception) & the vehicle can communicate with its dependent units
				& local & \si{100}{m} & \si{1}{Mbps} & \si{10}{ms} & sensor data\\
				
				2       & Parallel Driving                        & two or more vehicles of the same or different type communicate with each other
				& local & \si{100}{m} & \si{1}{Mbps} & \si{10}{ms} & coordinated driving\\
				
				3       & Local Control                           & without infrastructure any data in the vehicle can be accessed by an operator connected to it
				& local & \si{500}{m} & \si{1}{Gbps} & \si{1}{ms} & remote control, video\\
				
				4       & Data Offload to the Cloud               & in the presence of the ``infrastructure'' the vehicle off loads data
				& local + uplink & \si{100}{m} & \si{1}{Gbps} & \si{100}{ms} & bulk data\\
				
				5       & Remote Control                          & any data of the vehicle can be accessed by an operator globally
				& global & N/A & \si{100}{Mbps} & \si{10}{ms} & remote control, video\\
				\hline
		\end{tabular}}
	\end{table*}
	\begin{figure*}[t]
		\includegraphics[width=1.007\textwidth]{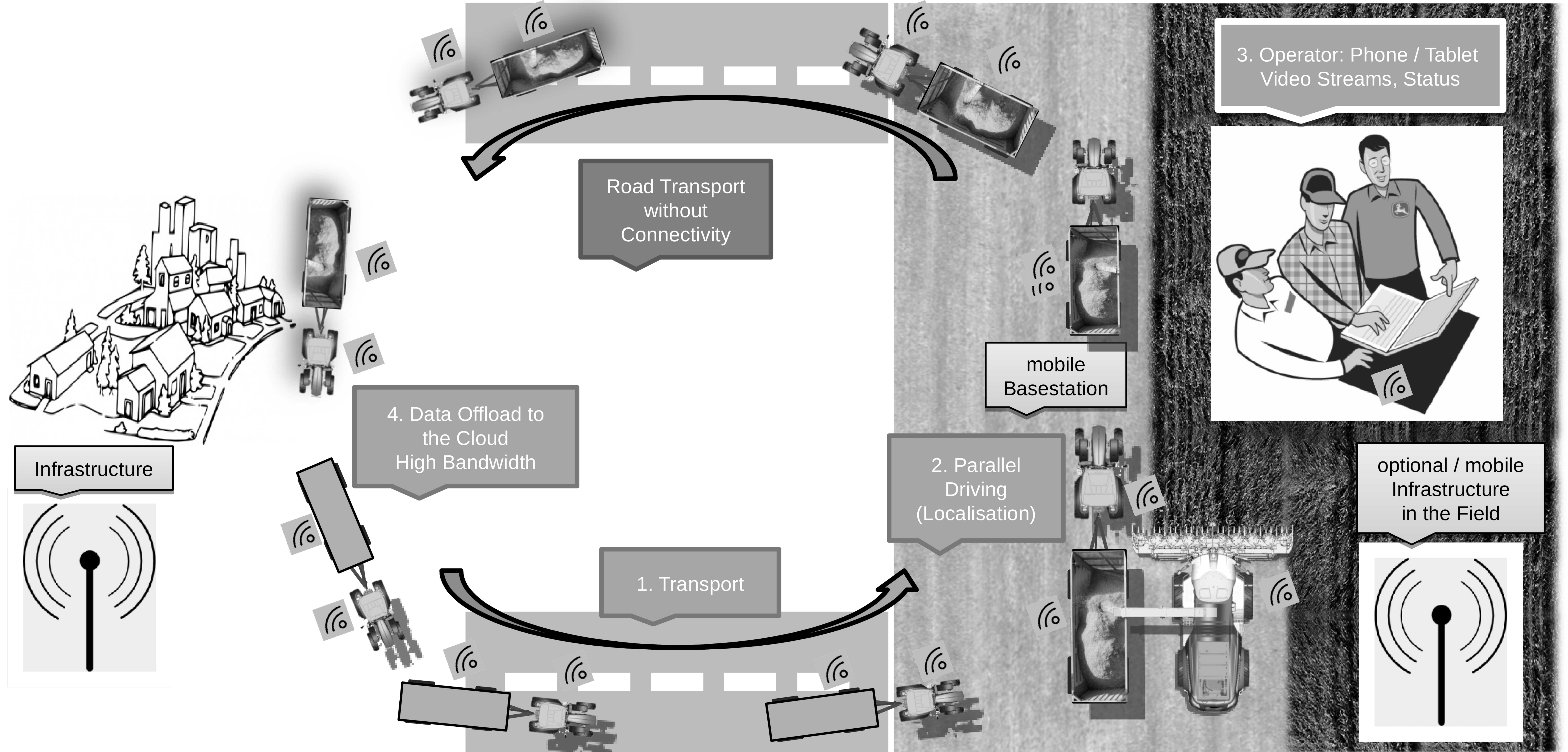}
		\caption{Illustration of Agricultural Use Case Class}\label{fig:agr-use-cases}
	\end{figure*}
	This class contains all the agricultural use cases to be explored in the AMMCOA project.
	In Tab.~\ref{tab:agr-use-cases} each individual use case is listed along with its description and requirements.
	The use case IDs correspond with the numbering in Fig.~\ref{fig:agr-use-cases} which shows a scenario wherein all but one of the individual use cases are cycled through during the course of the illustrated super use case.
	The cycle starts with the transport use case 1 (UC1) where the vehicle can be controlled remotely due to existing infrastructure connectivity.
	Upon arrival at the working area the vehicle connects to the local network and from there on it will be controlled by the machine which currently is designated as the coordinating vehicle (CV).
	The CV coordinates the parallel driving (UC2) until the transport capacity of the trailer is reached.
	At this point the vehicle leaves the working area and transports the load to the collection facility.
	Operators can check or remote control the vehicle (UC3) at the working area as shown in Fig.~\ref{fig:agr-use-cases}.
	Whenever the vehicle reaches an area with infrastructure connectivity, collected data can be pushed to the cloud (UC4) for further analyses and accounting.
	
	\subsection{Construction Use Case Class}
	\begin{table*}[t]
		\centering
		\caption{Construction Use Cases and Requirements}\label{tab:constr-use-cases}
		\resizebox{\textwidth}{!}{%
			\begin{tabular}{|l|p{5cm}|l|l|l|l|p{9cm}|}
				\hline  & & & & & & \\
				ID      & Use Case & Scope & Range & Throughput & latency & Application\\
				\hline  & & & & & & \\
				
				6       & Autonomous Compaction
				& local & \si{300}{m} & \si{1}{Gbps} & \si{10}{ms} & sensor data / autonomy planning: distances, maps, trajectories\\
				
				7       & Water Tanking
				& local & \si{5}{km} & \si{256}{kbps} & \si{1}{s} & rough positioning and status information\\
				
				8       & Construction Site Manager (onsite)
				& local & \si{100}{m} & \si{150}{Mbps} & \si{50}{ms} & monitoring / configuration\\
				
				9       & Construction Site Manager (offsite)
				& global & N/A & \si{150}{Mbps} & \si{1}{s} & monitoring\\
				
				%
				\hline
		\end{tabular}}
	\end{table*}
	In Tab.~\ref{tab:constr-use-cases} all construction uses cases considered in the project are listed and form a super use case to the one depicted in Fig.~\ref{fig:agr-use-cases}, where a vehicle, previously driving in formation, separates from the group and moves to the tanking vehicle to refill its water supply.
	Beginning with UC6, we have a paver followed by a series of rollers which need to coordinate with each other and  move back and forth. There is a planning computer mounted on the paver (CV) which needs to communicate with the rollers. The rollers also communicate with each other directly to improve localization and planning.
	The capacity of the roller's water tanks is usually not sufficient for a whole day operation. Therefore, the rollers need to refill their water supply (UC7). In order to do this autonomously the rollers need to get information about the tanking vehicle's position. Although the bandwidth requirements are quite low here, the required communication range (up to \si{5}{km}) is comparatively large.
	After refilling the roller returns to the construction site, which completes the super use cases' cycle.
	While being at a construction site, a site manager with a mobile device can communicate with the computer on the paver to get status information of the ongoing construction or to change settings (UC8), e.g. required number of passes for compaction.
	The back office on the other hand wants to get information of the current status of the construction site which constitutes UC9.
	
	There are situations, in both use case classes, where an individual vehicle changes between different modes of connectivity, like a switch from one RAT to another for example.
	The RAT switch could be caused by obstructed line of sight, a vehicle moving out of range of any particular RAT or generally bad propagation conditions due to environmental circumstances like heavy rain.
	This creates handover scenarios in both cases which have to be smooth and transparent.
	Participants should be notified about the currently available bandwidth and delays, so they can implement strategies for different circumstances.
	
	\section{Comparison of Network Architectures with autonomy support}
	In this section we give an overview of the considered concepts of network architectures that can operate independently from public network infrastructure. 
	Specifically, we will look at three related concepts, ``Trust Zone'', ``5G Islands'' and ``5G Private Networks''. In all cases, we add the functionality of moving network elements that are typically not allowed in mobile networks.
	In addition, we assume in all cases that certain network and control functions should be available locally after disconnection the AMMCOA network from the public network. One option is to use the functionalities supporting mobile edge clouds that already provide a certain degree of autonomy and thus resilience in cases where the backhaul connection is congested or otherwise unavailable.
	This is enabled by software defined networking (SDN) and network function virtualization (NFV) technology. Solutions for self management of SDN / NFV concepts already exists and allow self-optimisation and self-protection \cite{selfnet}.
	
	As described in \cite{7998270} a \textbf{Trust Zone (TZ)} is essentially defined as a set of network functions covering a certain geographical area via an edge cloud which serves a local set of base stations.
	The TZ autonomously implements different security policies while providing a certain set of services, like authentication, authorization and accounting (AAA), independently from the central cloud.\label{key}
	The main concern of a TZ is security.
	This concept does not differentiate between different security related virtual network functions (VNF) regarding their implementation and opportunity costs.
	All those functions are regarded as a whole and the decision to implement those in the local edge cloud is made globally depending on which costs are higher.
	
	The concepts of the \textbf{5G Island} generalizes the ideas of the TZ to other functions which are not security related and minimizes the sum of implementation costs (for local VNFs) and opportunity costs.
	If the opportunity costs for any given VNF are higher than its implementation costs then the respective function will be implemented locally on the edge cloud.
	This decision is made for each individual VNF.
	The opportunity costs are weighted and can change dynamically with the currently estimated outage probability of any given VNF.
	
	\textbf{5G private networks} are completely independent networks comprising core and access network as well as all required functions by themself. This implies, that it has its own home subscriber server (HSS) along with all the necessary user data in order to support its private users.
	Companies might use them to create their own cellular networks to e.g. establish a private, campus wide communication service. Private networks are currently finding significant interest in the vertical industries since the German national regulator has proposed to identify seperate spectrum for regional and local networks that could be used by private networks.

	\section{AMMCOA Network Architecture}
	
	\begin{figure*}[t]
		\includegraphics[width=\textwidth]{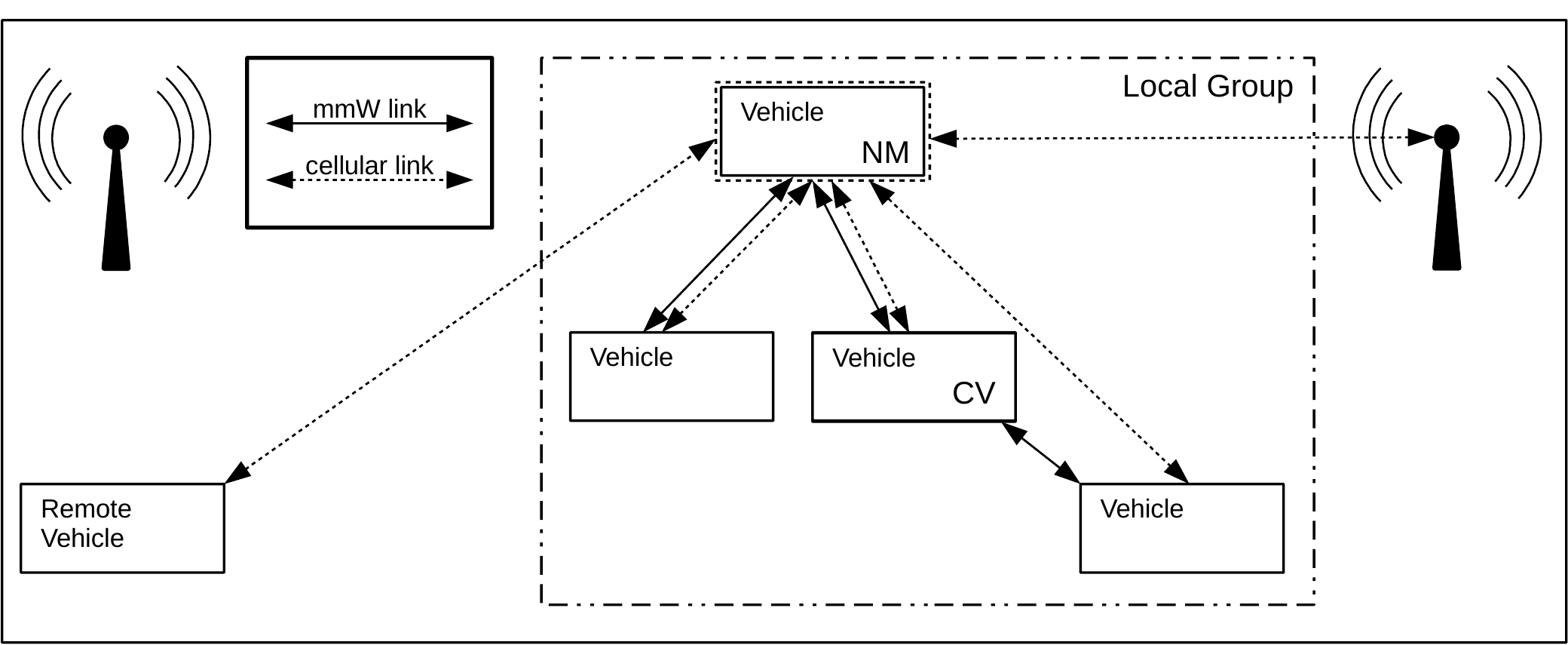}
		\caption{Illustration of the AMMCOA Network Architecture}\label{fig:net}
	\end{figure*}
	
	This section will give a brief overview of the envisioned AMMCOA network architecture.
	
	Of all the network concepts discussed in the previous section the ``5G Islands'' and ``5G Private Networks'' architectures are the best fit for the AMMCOA scenario.
	The main difference between these two is the ownership of the network.
	The former one is under the supervision of a particular mobile service provider whereas the later can be owned by either the fleet supplier or the fleet operator.
	In cases where it is desirable that the network is controlled by e.g. the fleet supplier, a derivative of the ``5G Private Network'' namely the ``Nomadic 5G Private Network'' looks like the logical choice.
	The major difference to consider is that this network is mobile and moves about.
	This means constantly changing propagation and connectivity conditions and challenges on the synchronisation of networks in cases of neighboured TDD networks.
	Here dynamic changes in the network topology have to be mastered by the above described context management.

	\subsection{Cellular Links}
	In each group one of the vehicles will be in the role of the network master (NM). Overall link reliability is for example achieved by multi connectivity solutions \cite{multiconnect}.
	All other vehicles of that group connect to it.
	See Fig.~\ref{fig:net} for reference.
	The NM functions as a base station whereas all other vehicles assume the role of a user equipment (UE).
	Furthermore, the NM vehicle and the coordinating vehicle (CV), don't need to be the same.
	These are separate concerns and roles.
	The CV is the one which will coordinate or control the other vehicles in order to work together.
	If the remote vehicle, as shown in Fig.~\ref{fig:net}, would get a considerably better connection to the infrastructure, it could make sense to switch its role to NM thereby providing the whole group a possibly higher bandwidth connection.
	The new NM would act as a gateway to the infrastructure and would be part of the local group from there on.
	In those kind of scenarios the local network hierarchy changes and a number of handover operations have to be performed without disrupting the throughput and latency performances.
	The connection to the central infrastructure is of course optional and the network has to be able to function without it.

	The high bandwidth short range communication will be implemented via point to point mmW links in the \si{26}{GHz} band.
	A set of directive antennas will be used to switch between communication partners.
	These links serve high data rate and ultra low latency applications like closed loop control driving.
	
	\section{Acknowledgements}
	Part of this work has been performed in the framework of the BMBF project AMMCOA. The authors would like to acknowledge the contributions of their colleagues, although the views expressed are those of the authors and do not necessarily represent the project.
	
	\bibliographystyle{IEEEtran}
	\bibliography{ref}

	
\end{document}